\documentclass[a4paper,11pt]{article}
\usepackage{pos}
\usepackage{textcomp}
\usepackage{inputenc}

\title{KM3NeT/ORCA: status and perspectives for neutrino oscillation and mass hierarchy measurements}
\author*[a]{Piotr Kalaczyński}

\affiliation[a]{National Centre for Nuclear Research,\\
Pasteura 7, Warsaw, Poland}

\emailAdd{piotr.kalaczynski@ncbj.gov.pl}
\abstract{
A next-generation neutrino telescope infrastructure, the Kilometer Cube Neutrino Telescope KM3NeT, is currently under construction in the Mediterranean Sea. Its low energy configuration ORCA is optimised for the detection of atmospheric neutrinos with energies above ∼1 GeV. The main goal of the ORCA detector is the precise measurement of atmospheric neutrino oscillation parameters and the determination of the neutrino mass ordering. The detector is also sensitive to a variety of other physics topics, such as dark matter, non-standard interactions and sterile neutrinos. An overview is presented of the ORCA detector and its research programme, along with early analyses of the data collected with the array in 4-lines configuration.
}

\FullConference{%
  40th International Conference on High Energy physics - ICHEP2020\\
  July 28 - August 6, 2020\\
  Prague, Czech Republic (virtual meeting)
}

\begin{document}

\maketitle

\section{Introduction and status}

KM3NeT is a distributed neutrino research infrastructure being constructed at the bottom of the Mediterranean Sea. It consists of two neutrino detectors: ARCA (Astroparticle Research with Cosmics in the Abyss) located off-shore Portopalo di Capo Passero, Sicily, Italy, at a depth of \(\sim\)3500\,m and ORCA (Oscillation Research with Cosmics in the Abyss) off-shore Toulon, France, at a depth of \(\sim\)2450\,m. The ORCA detector is designed to study the neutrino mass ordering (NMO) and other topics related to neutrino oscillations using atmospheric neutrinos in the GeV range \cite{LoI}.

The KM3NeT detectors consist of vertically aligned detection units (DUs), each carrying 18 digital optical modules (DOMs). Each DOM contains 31 3-inch photomultiplier tubes (PMTs), calibration and positioning instrumentation and readout electronics boards. ORCA has a horizontal spacing of \(\sim\)20\,m and vertical spacing of \(\sim\)9\,m between the DOMs, which is optimised for the search for low-energy neutrinos. 

The first DUs of the ORCA detector are already taking data since July 2019 and a number of
neutrino candidates have been observed \cite{Neutrino-flux}.

\section{Measurements}

Atmospheric muons are the main source of background for atmospheric neutrinos. They are much more abundant than neutrinos and have to be understood and suppressed in order to reliably select atmospheric neutrino candidate events. Data collected with the 4-DU configuration (ORCA4) from July 2019 to January 2020 have been used to measure the atmospheric muon rate and consequently, the atmospheric neutrino flux.

\subsection{Muon bundle rate}

\begin{figure}[h]
    \centering
    \includegraphics[width=0.49\textwidth]{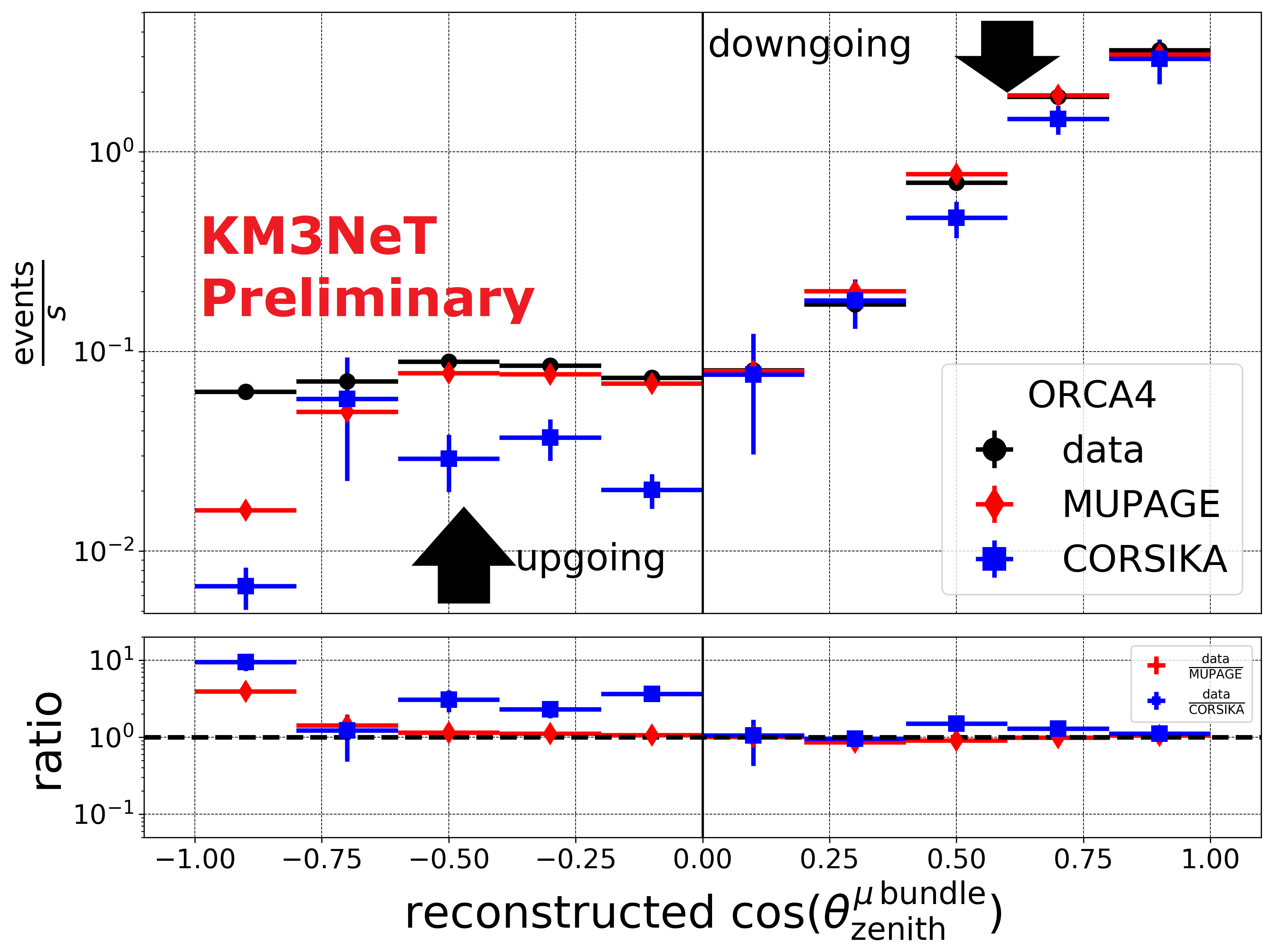}
    \includegraphics[width=0.49\textwidth]{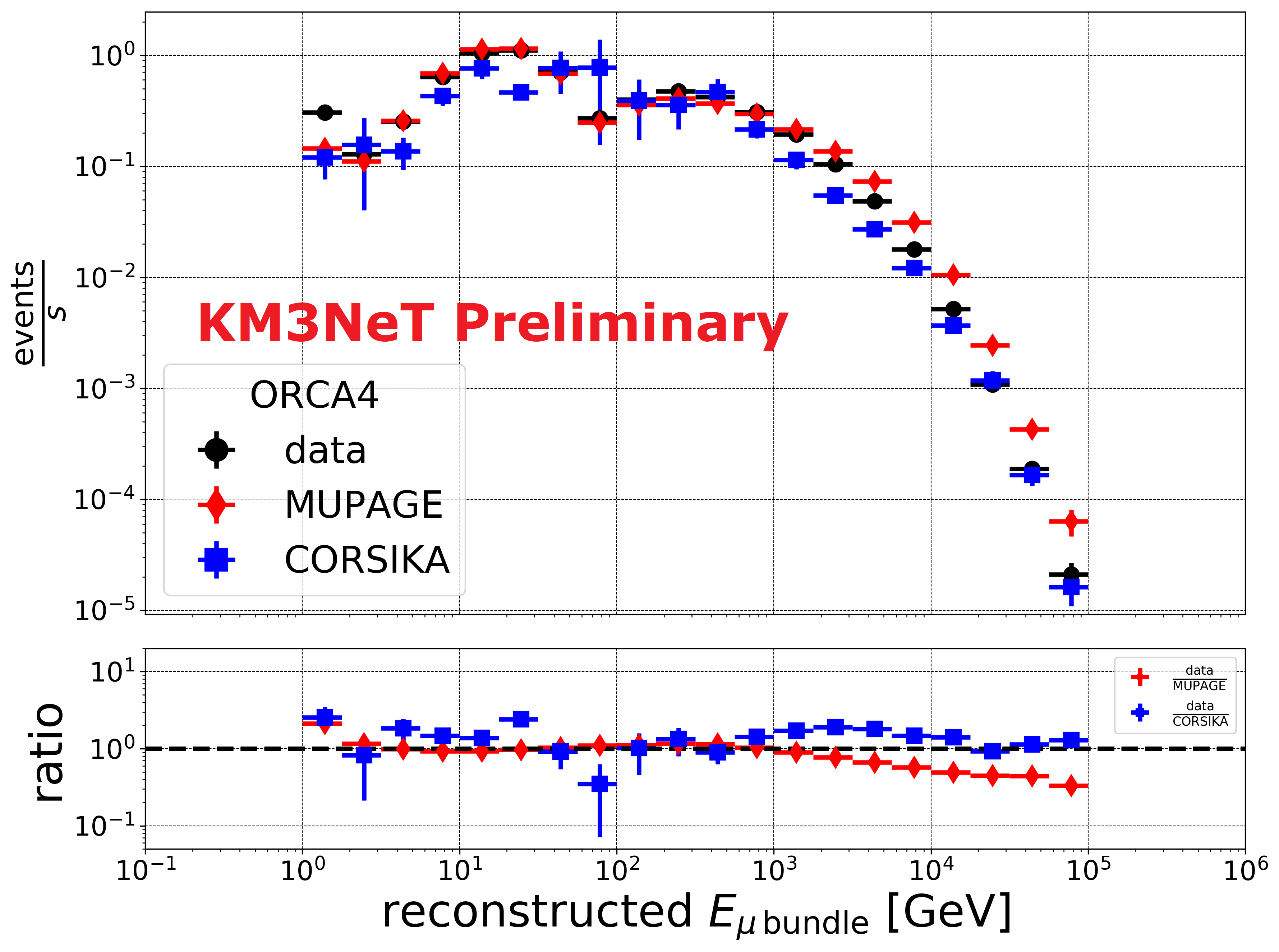}
    \caption{Atmospheric muon bundle rate as a function of the reconstructed zenith angle (left) and reconstructed bundle energy (right) for ORCA4 data \cite{Muon-rate}.}
    \label{fig:muon_rate}
\end{figure}

The rate of atmospheric muon bundles has been studied by comparing data corresponding to a livetime of about 35 days with MC simulations using the MUPAGE \cite{MUPAGE} and CORSIKA \cite{CORSIKA} programs . The comparison is based on the reconstructed zenith (\(\cos{\theta_{\mathrm{zenith}}}\)) and energy (\(E_{\mu \mathrm{\, bundle}}\)) distributions as shown in Figure \ref{fig:muon_rate}. The total muon rate was measured to be \(\sim\)455k \(\frac{\mu}{\mathrm{day}}\) \cite{Muon-rate}. From  Fig. \ref{fig:muon_rate} it is evident that the atmospheric muon contribution dominates over all zeniths and energies.

\subsection{Neutrino flux}

\begin{figure}
    \centering
    \includegraphics[width=0.99\textwidth]{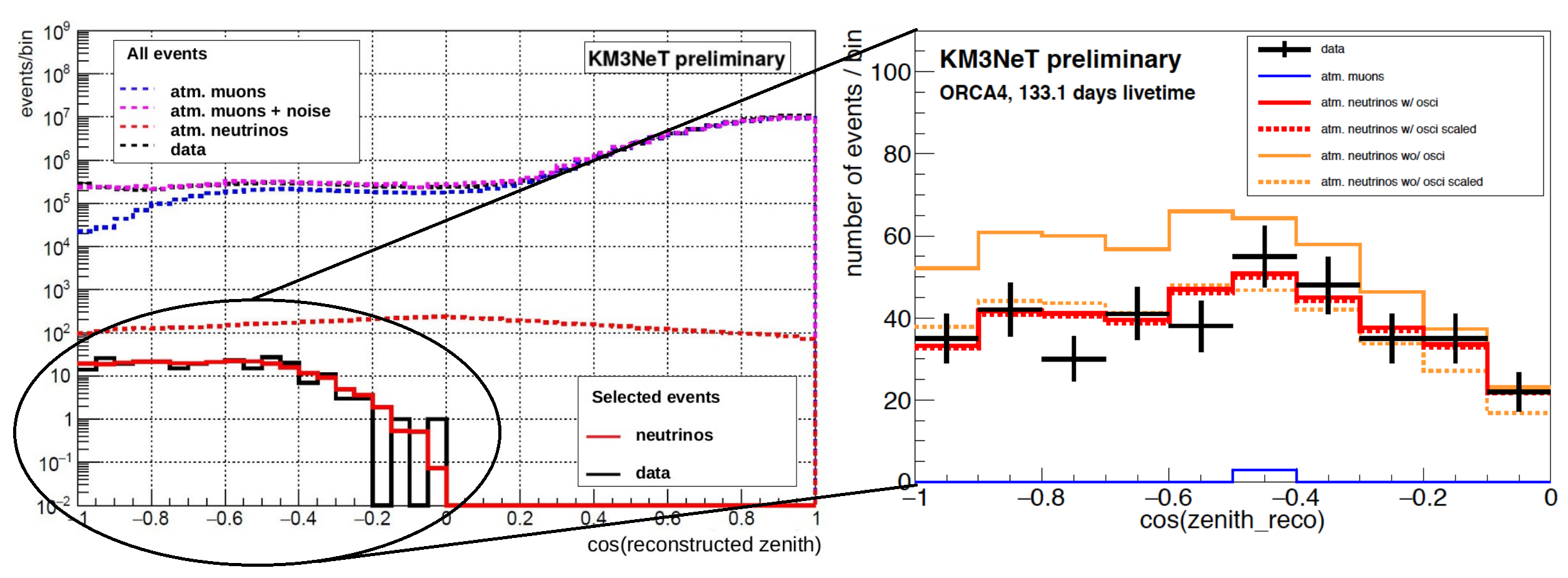}
    \caption{Left: Distribution of the reconstructed zenith angle for all reconstructed events (dotted lines) and for events surviving the selection criteria (solid lines) for ORCA4 data and MC simulated events. For selected events a  good agreement with the atmospheric neutrino MC is observed confirming that mostly neutrinos have been selected. Right: Comparison of the cosine of the reconstructed zenith angle with MC simulations for scenarios with and without neutrino oscillations \cite{Neutrino-flux}.}
    \label{fig:neutrino_flux}
\end{figure}

The atmospheric neutrino flux study uses data with a livetime of 4.5 months. A set of cuts is applied to select atmospheric neutrino events and the purity (expected percentage of neutrino events) of the resulting sample, shown in Figure \ref{fig:neutrino_flux}, is 99\% \cite{Neutrino-flux}. The observed rate is approximately 3 \(\nu\) per day. Even with the limited statistics offered by the data recorded with 4 ORCA DUs, comparison with MC simulation for scenarios with and without neutrino oscillations indicates that KM3NeT/ORCA data favour the oscillation hypothesis (\( p = 0.17\)).

\section{Sensitivity studies}

There has been a number of sensitivity studies conducted for KM3NeT/ORCA. Here, only the sensitivity to the Neutrino Mass Ordering and the neutrino mixing parameters, which are a subject of an upcoming paper are presented.

\subsection{Neutrino Mass Ordering}

\begin{figure}
    \centering
    \includegraphics[width=0.49\textwidth]{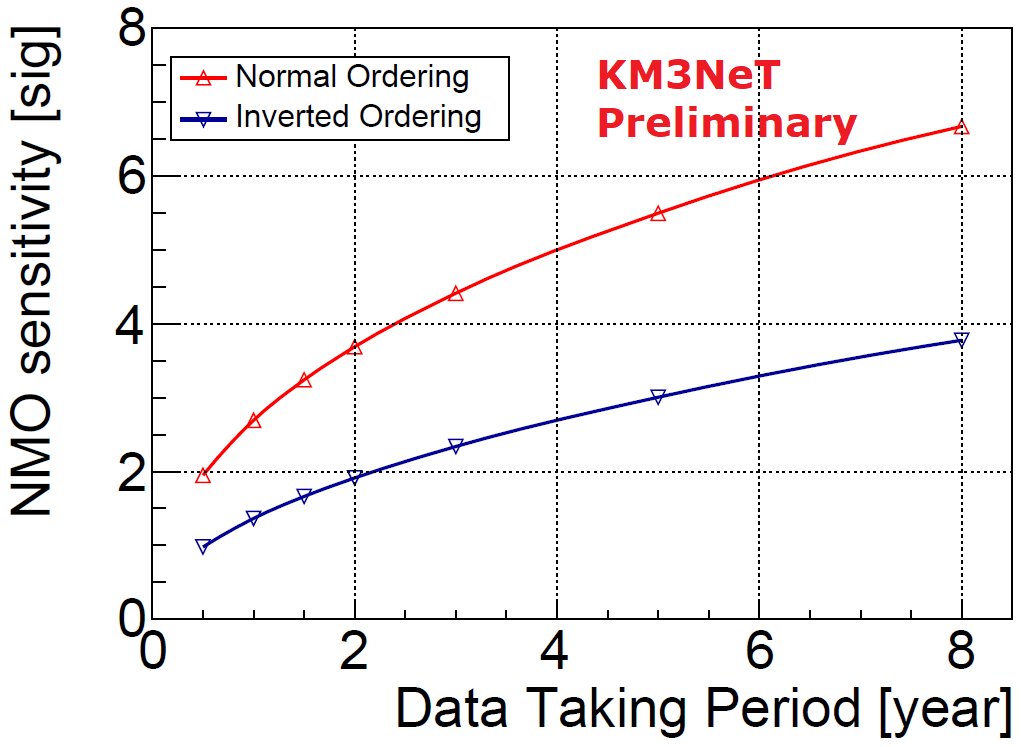}
    \includegraphics[width=0.49\textwidth]{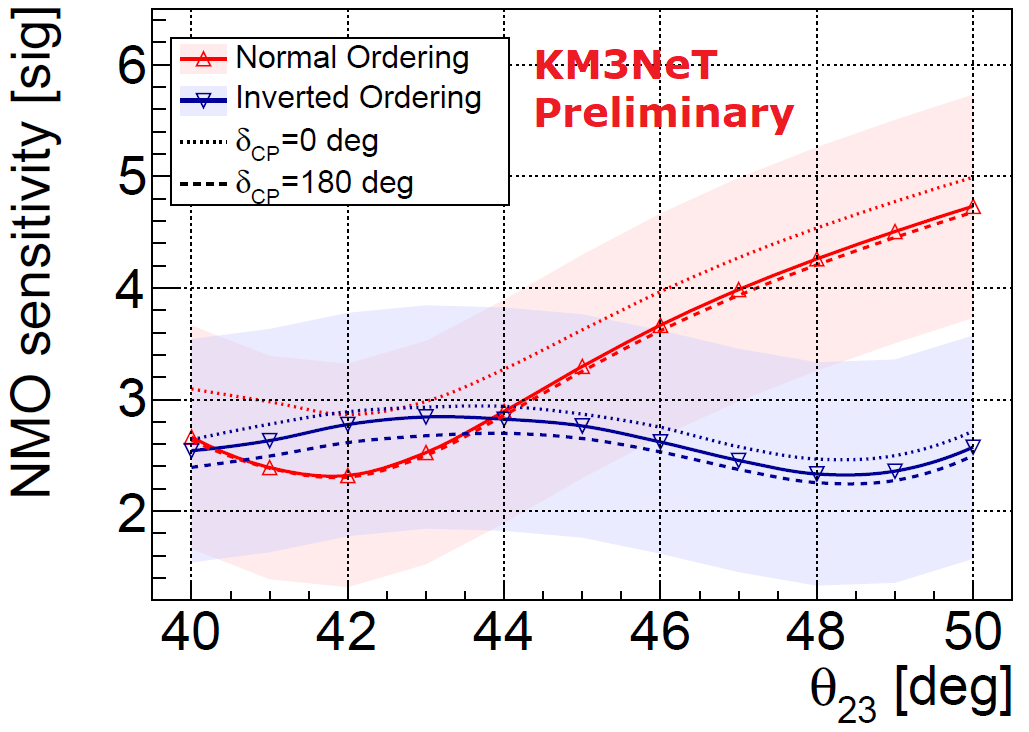}
    \caption{ORCA sensitivity to NMO as a function of the observation time in years (left plot) and \(\theta_{23}\) angle (right plot) for both NO and IO cases. For NO, the NMO could be determined with \(5\sigma\) after 4 years.}
    \label{fig:NMO_final_results}
\end{figure}

The NMO sensitivity is the flagship analysis for ORCA. The studies performed use atmospheric neutrino MC corresponding to a livetime of 3 years and to the full ORCA configuration with 115 DUs. Neutrino mixing parameter priors are taken from the three-flavour fit to the global neutrino oscillation measurements (NuFit 4.1) \cite{NuFIT}. The reconstruction of the vertex, direction, energy and time uses the maximum Likelihood approach. A set of reconstructed events is selected: only contained, upgoing events of high reconstruction quality (small \(\chi^2\)) are used. These selected events are then classified by the random decision forest algorithm into one of three classes. Most track events originate from \(\nu_{\mu}\) CC interactions, while events classified as showers are produced in \(\nu_{e}\) CC and \(\nu\) NC interactions. Afterwards, from the differences between the expected 2D distributions (in \(\cos{\theta_{\mathrm{zenith}}}\) and \(E\)) for the normal ordering (NO) and inverted ordering (IO), a sensitivity is computed by minimizing a test-statistic. By integrating the 2D sensitivities, one obtains the final result of the analysis, the expected sensitivity of ORCA to NMO, shown in Figure \ref{fig:NMO_final_results}.

%\begin{figure}
%    \centering
%    \includegraphics[width=0.99\textwidth]{NMO_3.PNG}
%    \caption{Classification of the neutrino events into track, middle and shower classes, shown for %each neutrino flavour separately for charged current (CC) interactions and for all neutrinos combined %for neutral current (NC) interactions.}
%    \label{fig:NMO_classes}
%\end{figure}

%\begin{figure}
%    \centering
%    \includegraphics[width=0.99\textwidth]{NMO_4.PNG}
%    \caption{2D distributions expected for the assumption of NO %(left column) and 2D sensitivities computed from them (right %column). The event classes are aligned from top to bottom: %track, middle, shower. The binning is 40x40.}
%    \label{fig:NMO_distributions}
%\end{figure}

\begin{figure}
    \centering
    \includegraphics[width=0.45\textwidth]{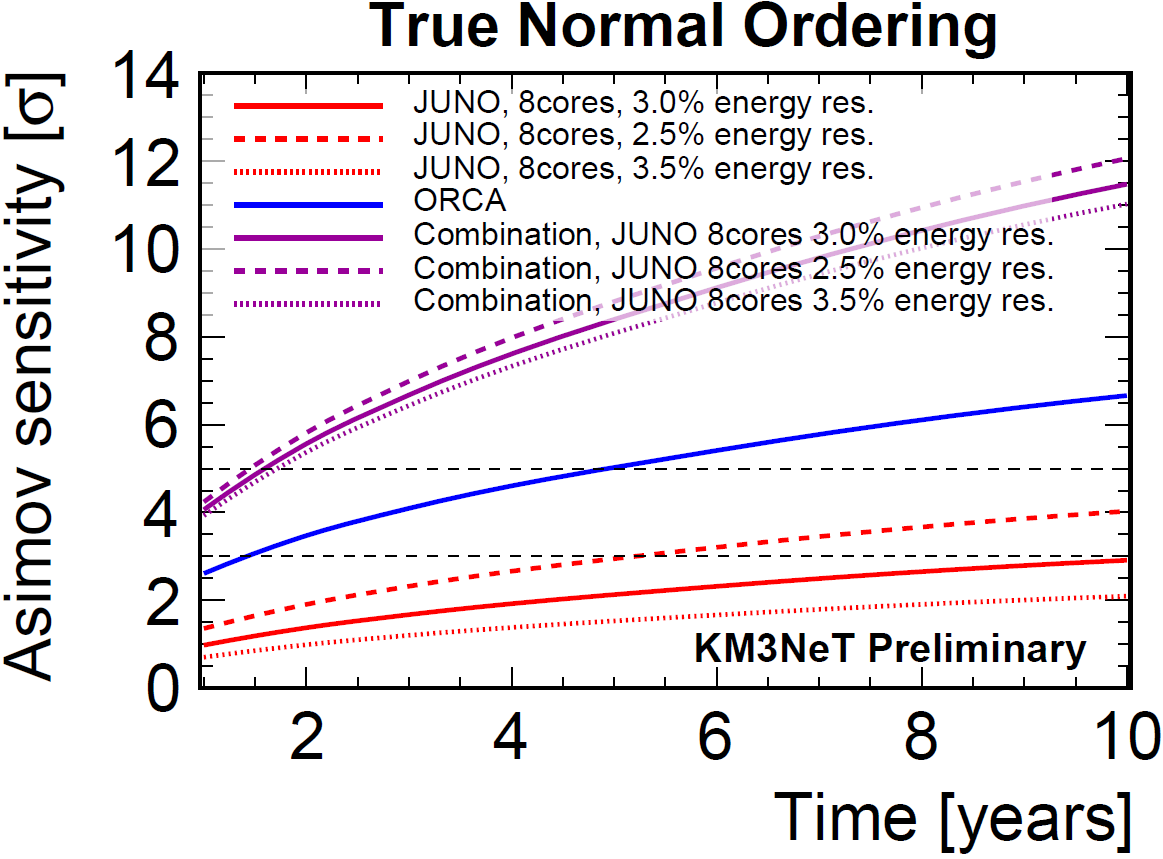}
    \includegraphics[width=0.45\textwidth]{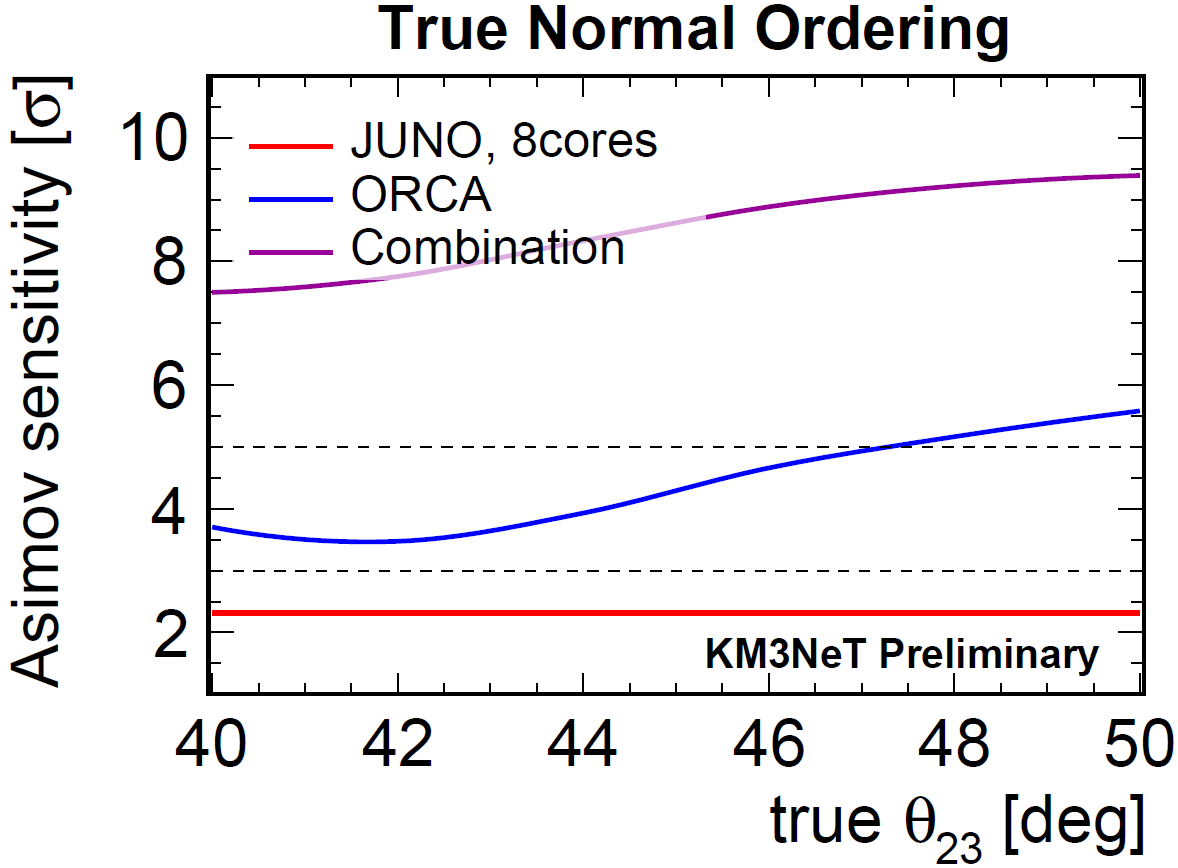}
    \caption{Sensitivity of ORCA+JUNO to NMO as a function of time (left plot) and \(\theta_{23}\) angle (right plot) for NO. From the combined analysis of ORCA and JUNO data the NMO could be determined with \(5\sigma\) after 1 year \cite{JUNO_sens}.}
    \label{fig:JUNO}
\end{figure}

There is also an independent study on the combined sensitivity of ORCA with JUNO \cite{JUNO}, using the same MC sample, but with slightly older priors (NuFit 4.0 \cite{NuFIT}) and a different method: \(\chi^2\) minimization of an Asimov dataset . The resulting sensitivity is shown in Figure \ref{fig:JUNO}.

\subsection{Neutrino mixing parameters}

\begin{figure}
    \centering
    \includegraphics[width=0.49\textwidth]{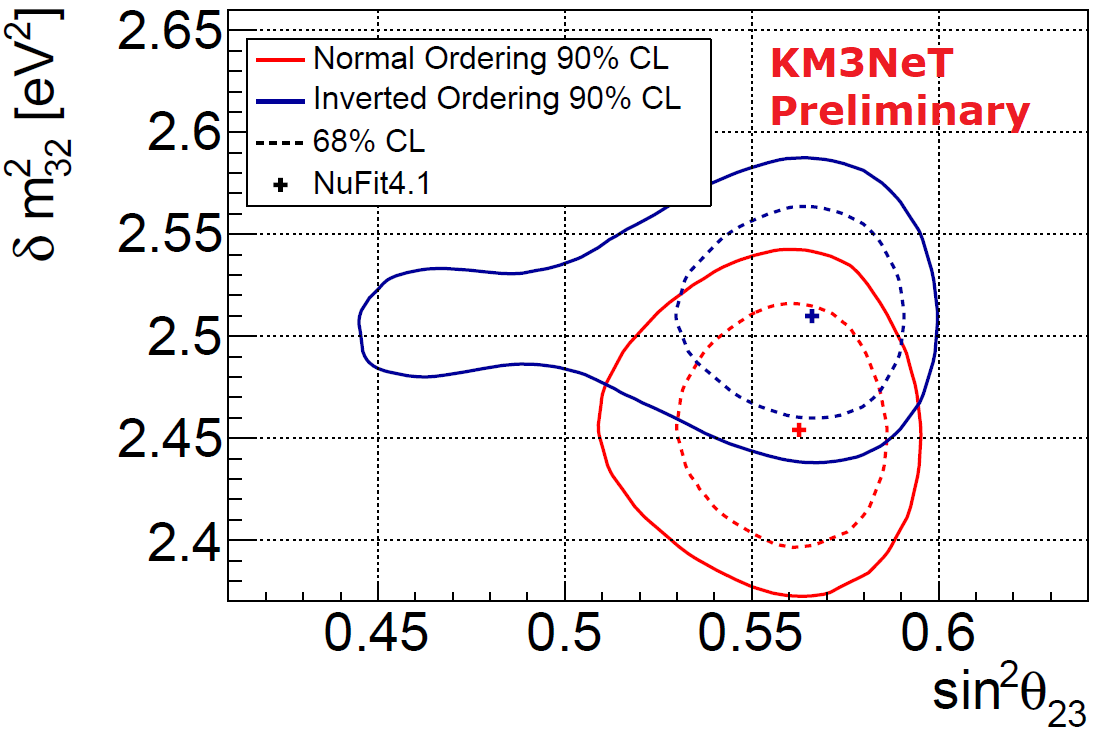}
    \includegraphics[width=0.49\textwidth]{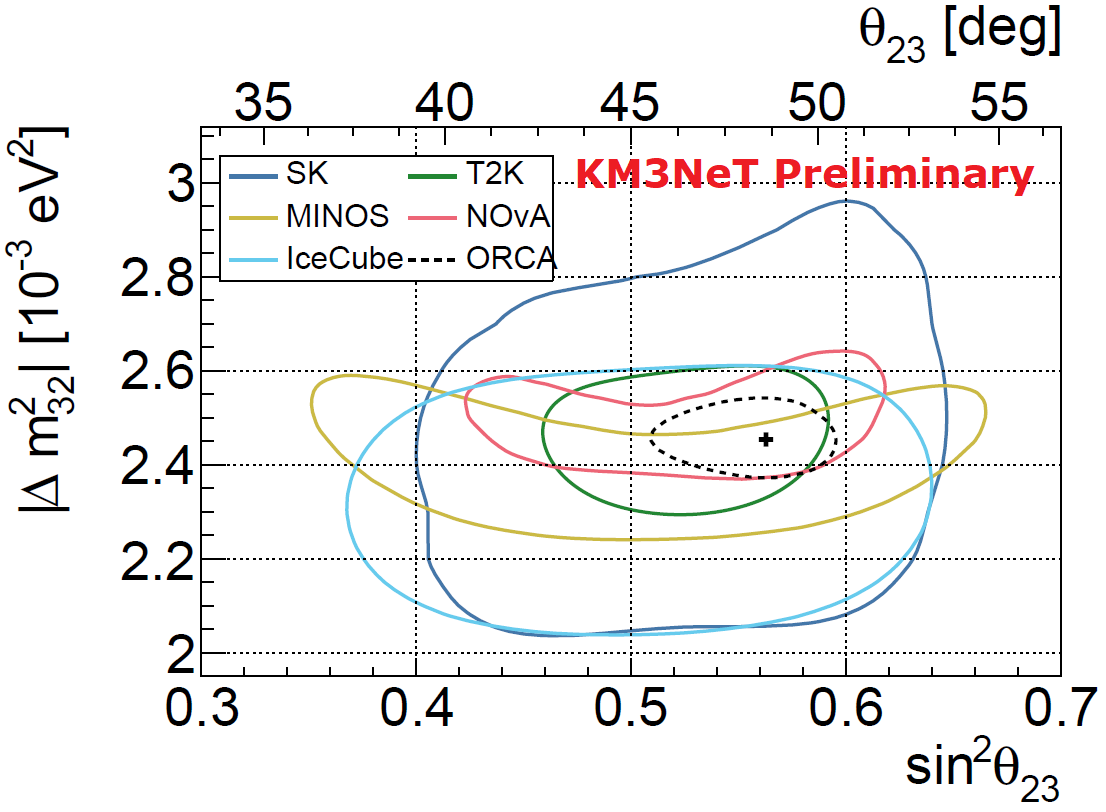}
    \caption{Sensitivity of ORCA to the values of \( \Delta m_{32}^2 \, [\mathrm{eV}^2] \) and \( \sin{\theta_{23}}^2 \) for NO and IO and in comparison with other experiments (assuming NO) \cite{IceCube,SK,T2K,MINOS,NOvA}.}
    \label{fig:mixing_params}
\end{figure}

Sensitivity of ORCA to the mixing parameters has been estimated using the same MC and method as for the NMO study. As can be seen in Figure \ref{fig:mixing_params}, KM3NeT/ORCA will offer a remarkable improvement in precision of measuring the neutrino mixing parameters, outperforming the existing neutrino experiments taking data for many years now \cite{IceCube,SK,T2K,MINOS,NOvA}.

%\subsection{\(\nu_{\tau}\) appearance}

%\subsection{Sterile neutrinos}

%\subsection{Non-standard interactions (NSI)}

%\subsection{Core-collapse Supernovae (CCSN)}

%\subsection{Dark matter from the Sun}

\section{Summary}

There are various analyses ongoing using data collected with 6 ORCA DUs (ORCA6) as well as sensitivity studies for the complete detector, which did not fit within the page limit of these proceedings, like \(\nu_{\tau}\) appearance \cite{Tau}, sterile neutrinos \cite{Sterile}, non-standard interactions \cite{NSI}, core-collapse supernovae \cite{CCSN} and dark matter. ORCA is running stably and is expected to be further expanded in 2020. First measurements performed with ORCA4, already indicate that it can observe the neutrino oscillations.  Some detector design improvements are also under discussion, such as Super-ORCA \cite{Super-ORCA} and P2O \cite{P2O}.

\vspace{0.5cm}

This work is supported by the National Science Centre, Poland (grant 2015/18/E/ST2/00758)

\end{document}